\documentclass[aps,prb,twocolumn,superscriptaddress,floatfix]{revtex4}
\usepackage{color}
\usepackage{graphicx}
\usepackage{epsfig}
\usepackage{textcmds}
\bibstyle{apsrev.bib}

\begin{document}
\input epsf.sty

\title{Density of critical clusters in strips of strongly disordered systems}

\author{M. Karsai}
\affiliation{
Institute of Theoretical Physics,
Szeged University, H-6720 Szeged, Hungary}
\affiliation{Institut N\'eel-MCBT
 CNRS 
\thanks{U.P.R. 5001 du CNRS, Laboratoire conventionn\'e
avec l'Universit\'e Joseph Fourier}, B. P. 166, F-38042 Grenoble,
France}
\author{I. A. Kov\'acs}
\affiliation{
Research Institute for Solid State Physics and Optics,
H-1525 Budapest, P.O.Box 49, Hungary}
\affiliation{Department of Physics, Lor\'and E\"otv\"os University, H-1117 Budapest,
P\'azm\'any P. s. 1/A, Hungary}
\author{J-Ch. Angl\`es d'Auriac}
\affiliation{Institut N\'eel-MCBT
 CNRS 
\thanks{U.P.R. 5001 du CNRS, Laboratoire conventionn\'e
avec l'Universit\'e Joseph Fourier}, B. P. 166, F-38042 Grenoble,
France}
\author{F. Igl\'oi}
\affiliation{
Research Institute for Solid State Physics and Optics,
H-1525 Budapest, P.O.Box 49, Hungary}
\affiliation{
Institute of Theoretical Physics,
Szeged University, H-6720 Szeged, Hungary}

\date{\today}

\begin{abstract}
We consider two models with disorder dominated critical points and study the distribution of clusters
which are confined in strips and touch one or both boundaries. For the classical random bond Potts
model in the large-$q$ limit we study
optimal Fortuin-Kasteleyn clusters by combinatorial optimization algorithm. For the
random transverse-field Ising chain clusters are defined and calculated through the
strong disorder renormalization group method. The numerically calculated density profiles close
to the boundaries are shown to follow scaling predictions. For the random bond Potts
model we have obtained accurate numerical estimates for the critical exponents and demonstrated
that the density profiles are well described by conformal formulae.
\end{abstract}

\maketitle

\newcommand{\bc}{\begin{center}}
\newcommand{\ec}{\end{center}}
\newcommand{\be}{\begin{equation}}
\newcommand{\ee}{\end{equation}}
\newcommand{\beqn}{\begin{eqnarray}}
\newcommand{\eeqn}{\end{eqnarray}}

\section{Introduction}

In a critical system the correlation length is divergent and correlated
domains appear in all length scales. This phenomenon is seen for percolation\cite{staufferaharony}
where the domains are connected clusters. In discrete spin models, such as in the Ising and
the Potts models, domains of correlated spins can be identified in different ways.
One possibility is to use geometrical clusters\cite{droplet} (also called Ising or Potts clusters)
which are domains of parallel spins. In two dimensions (2d) geometrical clusters percolate the sample
at the critical temperature and their fractal dimension can be obtained through
conformal invariance\cite{2dgeom}.
This value is generally different from the fractal dimension of Fortuin-Kasteleyn (FK) clusters\cite{kasteleyn}
which are represented by graphs of the high-temperature expansion. From a geometrical
cluster the FK cluster is obtained by removing bonds by a probability, $1-p=e^{-K_c}$,
$K_c$ being the critical value of the coupling. The fractal dimension of a FK cluster
is directly related to the scaling dimension of the magnetization.

In a finite geometry, such as inside strips or squares, one is interested in the
spanning probability and different crossing problems of the critical clusters. For
2d percolation many exact and numerical results have been obtained in this field\cite{Cardy92,Ziff92,LanglandsEtAl92,MA,KZi,S01,BD,KlebanZagier03,JD06}.
Another interesting problem is the density of clusters in restricted geometries\cite{SKDZ07}, 
which is defined by the fraction of samples for which a given point belongs to a cluster with
some prescribed property, such as touching the edges of infinite and half-infinite strips, squares, etc.
This latter problem is analogous to the calculation of order parameter profiles in restricted
geometries, which has been intensively studied through conformal invariance and numerical methods\cite{BE85,CFT87a,CFT87b,CFT90,CFT91a,CFT91b,CFT94,CFT97a,CFT97b,CFT98,ResStraley00,CFT00a,CFT00b,CFT00c,CFT01}.

Correlated clusters can be defined also in models in the presence of quenched disorder. In a
random fixed point one generally considers such quantities, which are averaged first over
thermal fluctuations and afterwards over quenched disorder. In isotropic random systems conformal
symmetry is expected to hold at the critical point so that average operator profiles
and average cluster densities are expected to be invariant under conformal transformations.
Among disordered systems an interesting class is represented by such models in which the
transition in the pure version is of first-order, but in the disordered version the transition
softens into second order\cite{Cardy99}. This type of random fixed point can be found, among
others, in the two-dimensional random bond Potts model (RBPM) for $q>4$, $q$ being the number of states\cite{pottsmc,pottstm}.

If the distribution of the disorder is not isotropic, e.g. it has a layered structure,
then the scaling behavior of the disordered system can be anisotropic, which is manifested
in the fact that the critical clusters have an elongated shape. This means that the characteristic sizes
of the clusters parallel, $\xi_{\parallel}$, and perpendicular, $\xi_{\perp}$, to the layers are
generally related as $\xi_{\parallel} \sim \xi_{\perp}^{z}$, with an anisotropy exponent, $z \ne 1$.
These essentially anisotropic models are not conformally invariant.
A well known example in this class is the McCoy-Wu model\cite{mccoywu}, which is a two-dimensional
Ising model with layered disorder. Study of this system, as well as its one-dimensional quantum
version the random transverse-field Ising chain (RTFIC) has shown\cite{fisher} that the critical behavior
is controlled by a so called infinite disorder fixed point (IDFP), in which scaling
is strongly anisotropic\cite{im}. The characteristic lengths are related as $\ln \xi_{\parallel} \sim \xi_{\perp}^2$ so that the anisotropy exponent is formally infinite. The same IDFP is found
to control the
critical behavior of the randomly layered $q$-state Potts model\cite{SM}, as well as for strong enough
layered disorder the critical behavior of percolation\cite{ji02} and directed percolation\cite{hiv03}. Operator profiles in the RTFIC have
been studied numerically\cite{CFT97b} and
the obtained data could be well fitted by curves which are obtained by analogy of the
conformal results.

In this paper we study the density of critical clusters in two problems in which the critical properties
are dominated by strong disorder effects. The first model is the two-dimensional RBPM in the
large-$q$ limit. In this model for a given realization of disorder the high-temperature series expansion is dominated by a single graph\cite{JRI01}, the so called optimal diagram, thus thermal fluctuations
are indeed negligible.
This optimal diagram is calculated for each finite sample by a combinatorial
optimization algorithm\cite{aips02}.
Clusters in the optimal diagram are isotropic and density of clusters is obtained through averaging
over disorder realizations. The second model we consider is the RTFIC, i.e. a random quantum model
which is related to the classical McCoy-Wu model, in which the Fortuin-Kasteleyn clusters are
strongly anisotropic. In the RTFIC clusters of correlated spins can be defined and calculated by
the so called strong disorder renormalization group (SDRG) method\cite{im}. During renormalization the
system is transformed into a set of effective spin clusters and for a finite system with a
given realization of the disorder one obtains the final cluster, which contains the mostly correlated sites.
The fractal dimension of the final cluster at the critical point is directly related to the
scaling dimension of the magnetization of the RTFIC. Here we calculate the density of these final
clusters which are confined in a (one-dimensional) strip.

The two models we study in this paper are expected to be closely related, as far as
their critical properties are concerned. Based on numerical and analytical studies\cite{ai03,mai04} the
scaling dimension of the magnetization, $x_b$, and that of the surface magnetization, $x_s$,
are conjectured to be the same for both systems and given by\cite{fisher}:
\be
x_b=\frac{3-\sqrt{5}}{4},\quad x_s=\frac{1}{2}\;.
\label{x_x_s}
\ee
On the other hand the correlation length exponents are related by a factor of two: it is $\nu=2$ for
the RTFIC and $\nu=1$ for the RBPM. Here we are interested in a possible analogy
in terms of the densities of the critical cluster.

The structure of the paper is the following. Sec.\ref{sec:RBPM} is devoted to the RBPM. Here
we define the model, outline the calculation of the optimal diagram and then analyze the
statistics of the distribution of the clusters. The numerically calculated densities are
then compared with formulae which are obtained by modifying conformal results for percolation.
In Sec.\ref{sec:RTFIC} we define the RTFIC, recapitulate the essence of the SDRG method and
then numerically calculate final clusters at the critical point. The numerically calculated
densities are compared with analytical formulae in this case, too. The paper is closed with
a discussion.

\section{Optimal clusters in the RBPM}
\label{sec:RBPM}

\subsection{Model}
The $q$-state Potts model\cite{Wu} is defined by the Hamiltonian:
\begin{equation}
\mathcal{H}=-\sum_{\left\langle i,j\right\rangle }J_{ij}\delta(\sigma_{i},\sigma_{j})
\label{eq:hamilton}
\end{equation}
in terms of the Potts-spin variables, $\sigma_{i}=0,1,\cdots,q-1$, at site $i$. The summation runs over all edges of a lattice $\langle i,j\rangle \in E$, and in our study the couplings, $J_{ij}>0$, are independent
and identically distributed random numbers. To write the partition sum of the system it is
convenient to use the random cluster representation\cite{kasteleyn}:
\begin{equation}
Z =\sum_{G\subseteq E}q^{c(G)}\prod_{ij\in G}\left[q^{\beta J_{ij}}-1\right]
\label{eq:kasfor}
\end{equation}
where $\beta=1/(k_B T \ln q)$, the sum runs over all subset of bonds, $G\subseteq E$ and $c(G)$ stands for the number of connected components of $G$. In the following we restrict ourselves to the square lattice,
in which case the phase transition in the non-random model is of second order (first order) for
$q \le 4$ ($q > 4$)\cite{baxter}, but for random couplings the phase transition softens to second order for
any value of $q$\cite{aizenmanwehr,imrywortis}. For conceptional simplicity we consider the large-$q$ limit, where $q^{\beta J_{ij}} \gg 1$, and the partition function can be written as
\begin{equation}
Z=\sum_{G\subseteq E}q^{\phi(G)},\quad \phi(G)=c(G) + \beta\sum_{ij\in G} J_{ij}\label{eq:kasfor1}
\end{equation}
which is dominated by the largest term, $\phi^*=\max_G \phi(G)$. Consequently for a given realization
of disorder the thermal fluctuations play a completely negligible role and the critical properties
of the system are dominated by disorder effects. The optimal diagram of the RBPM plays a completely
analogous role as the geometrical clusters in percolation theory. For example at the critical
point there is a giant cluster in the optimal diagram the fractal dimension
of which, $d_f$, is related to the scaling dimension of the (average) magnetization as $d=d_f+x_b$,
where $d=2$ is the dimension of the system. One can also study other questions, such as distribution
of the mass of the connected clusters, spanning probability, surface scaling exponent, etc. Here
we are going to investigate the density of clusters in strip geometry.

\begin{figure}
  \begin{center}
\includegraphics[width=3.35in,angle=0]{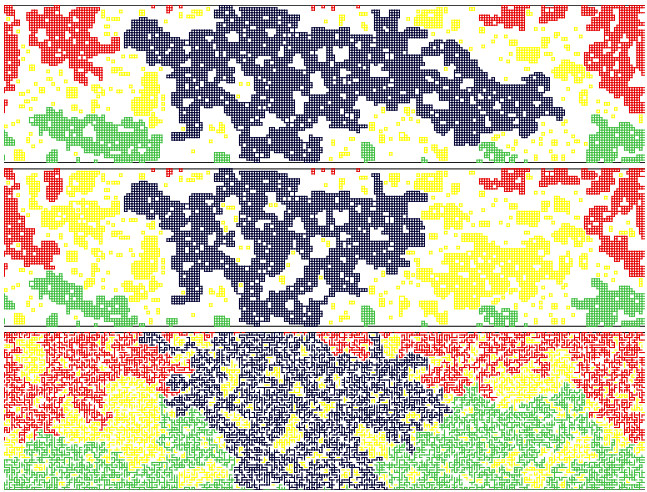}
   \end{center}
   \caption{(Color online) Different type of clusters in the optimal diagram of the RBPM:
spanning clusters [black],
clusters which touch only the upper (lower) boundary of the strip [red (green)] and clusters which have
no common points with the boundary [yellow]. Lower panel, $\Delta=1/2$, standard bond percolation;
middle panel, $\Delta=5/12$; upper panel, $\Delta=4/12$. Note that the breaking-up length is increasing
with decreasing $\Delta$.
}
   \label{fig:cluster}
 \end{figure}

During our study we use a bimodal form of the disorder, when the reduced couplings, $K_{ij}=\beta J_{ij}$
take two values: $K_1=K-\Delta$ and $K_2=K+\Delta$ with equal probability. Generally we study the
critical point of the system which is located at $K=K_c=1/2$\cite{dom_kinz} independently of the value of
$0 \le \Delta \le 1/2$. Note that the pure system is obtained for $\Delta=0$, whereas for $\Delta=1/2$,
when just the strong bonds are present in the system we have the traditional percolation problem.
The evaluation of the optimal diagram with decreasing values of $\Delta$ is shown in Fig.\ref{fig:cluster}.
Here one can see that with decreasing $\Delta$ the clusters become more compact. More precisely
one can define a finite length-scale, the so called breaking-up length, $l_b$, which is
rapidly increasing with decreasing $\Delta$. For small $\Delta$ the breaking-up length
has been calculated in Ref.\cite{mai04}:
\begin{equation}
l_b \approx l_0 \exp\left[ A \left( \frac{K}{\Delta} \right)^2 \right]\;.
\label{eq:l_b}
\end{equation}
which is divergent for $\Delta \to 0$, i.e. in the non-random system limit.
In a numerical calculation on a finite sample of linear size, $L$, one should have the
relation, $L \gg l_b$, thus $\Delta$ should be not too small. On the other hand one should
also be sufficiently far from the percolation limit, $\Delta=1/2$, in order to get rid of cross-over effects.
This means that the optimal choice of $\Delta$ is a result of a compromise, which
in our case seems to be around $\Delta=5/12$, when the
typical breaking-up length is about $l_b \sim 14$. Most of
our studies are made for this value, but in order to check universality, i.e. disorder independence of
the results we have made also a few calculations for $\Delta=21/48$, too.

Calculation of the optimal diagram for a given realization of disorder is a non-trivial
optimization problem, for which very efficient combinatorial optimization algorithm have
been developed\cite{aips02}, which works in strongly polynomial time. Application of this method
made it possible to obtain the exact optimal diagram for comperatively large finite systems.
In order to have an effective
strip geometry we have considered lattices of rectangle shape with an aspect ratio of four. The
strips have open boundaries along the long direction and periodic boundary condition was used
in the other direction. We
mention that the same geometry has been used before for percolation, too\cite{SKDZ07}. The width
of the lattices we considered are from $L=32$ up to $256$. Typically we have considered
several thousand samples, for the largest system we have thousand
samples.

\subsection{Densities of critical clusters}

We start to study the density of crossing clusters, $\rho_b(l/L)$, which is given by the probability
that a point in the position, $l$, measured perpendicular to the strip, belongs to a cluster which
touches both boundaries of the strip.
For percolation in the continuum limit, $l \gg 1$, $L \gg 1$ and $y=l/L$ the
density, $\rho_b(y)$, is calculated trough conformal invariance\cite{SKDZ07}:
\be
\rho_b(y) \propto \left(\sin \pi y\right)^{-x_b} \left[\left(\cos \frac{\pi y}{2} \right)^{x_s}+
\left(\sin \frac{\pi y}{2} \right)^{x_s}-1\right]\;
\label{rho_b}
\ee
in which $x_b=5/48$ and $x_s=1/3$ are the scaling dimensions for percolation\cite{staufferaharony}.

\begin{figure}
  \begin{center}
     \includegraphics[width=3.35in,angle=0]{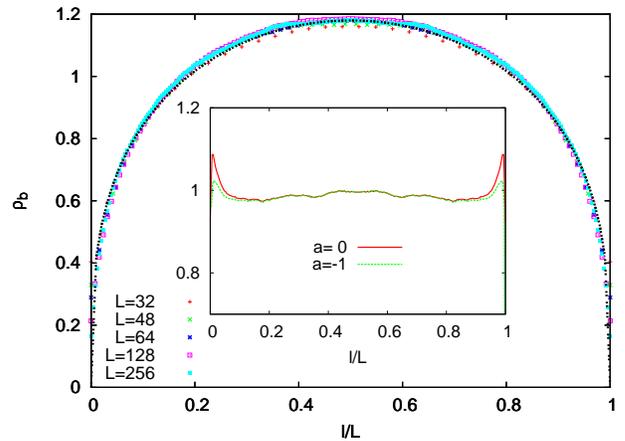}
   \end{center}
   \caption{(Color online) Normalized density profiles, $\rho_b(l/L)$, of the RBPM for different
widths. The dashed line indicates the conformal result in Eq.(\ref{rho_b})
with the conjectured exponents in Eq.(\ref{x_x_s}) and with the boundary parameter $a=0$ in Eq.(\ref{y_a}).
In the inset the ratio of simulation to theoretical
results are presented for $L=256$ and for two different boundary parameters: $a=0$ and $a=-1.0$,
see Eq.(\ref{y_a}).
}
   \label{fig:rho_b}
 \end{figure}

For the RBPM the numerically calculated normalized densities, $\rho_b(l/L)$, for different widths
are shown in Fig.\ref{fig:rho_b}.
All the data fit to the same curve and the finite breaking-up length, $l_b$, seems to have only
a small effect.

\begin{figure}
  \begin{center}
     \includegraphics[width=3.35in,angle=0]{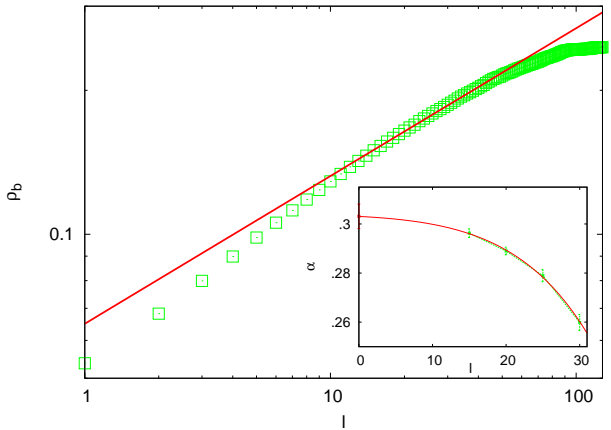}
   \end{center}
   \caption{(Color online) Density profile, $\rho_b(l)$, of the RBPM for $L=256$ close to the surface in
log-log plot. The straight (red) line
has a slope $x_s-x_b=0.309$. Inset: estimates of the slope using different windows of the fit, see the text.
Here the full (red) line indicates a parabolic fit.}
   \label{fig:rho_b_1}
 \end{figure}

In the surface region, $l \ll L$, but $l > l_b$ one expects from scaling theory: $\rho_b(l) \sim l^{x_s-x_b}$,
which is in accordance with the limiting behavior of the conformal prediction in Eq.(\ref{rho_b})
with the conjectured scaling exponents for $x_b$ and $x_s$ in Eq.(\ref{x_x_s}). In Fig.\ref{fig:rho_b_1}
we have presented $\rho_b(l)$ in a log-log plot in the surface region for the largest finite system.
Indeed, for $l>l_b$ the points are well on a straight line the slope of which is compatible with
the conjectured value: $x_s-x_b=0.309$. We have also estimated the asymptotic slope of the curve
by drawing a straight line through the points in a window $[l_b+l/2,l_b+3l/2]$ by least square fit.
Fixing $l_b=15$ the estimates with varying $l$ seem to have a $\sim l^2$ correction (see the inset
of Fig.\ref{fig:rho_b_1}) and the extrapolated slope is $x_s-x_b=0.303(8)$ in agreement with
the conjectured values in Eq.(\ref{x_x_s}).

We have also checked if the conformal result for percolation in Eq.(\ref{rho_b}) using the
conjectured scaling exponents for $x_b$ and $x_s$ in Eq.(\ref{x_x_s}) can be used to fit the scaling
curve for the RBPM for the whole profile. As seen in Fig.\ref{fig:rho_b} the agreement between the numerical results and the formula in Eq.(\ref{rho_b}) is indeed very good. We have also calculated the ratio of the simulation to the
theoretical results. In this case for the theoretical curve we used the variable: 
\be
y=(l+a)/(L+2a)\;,
\label{y_a}
\ee
where
$a=O(1)$ measures the effective position of the boundary in the lattice model. By varying $a$ one can obtain
a better fit in the boundary region, as seen in the inset of Fig.\ref{fig:rho_b}.

Next we consider the density of those clusters which are touching one boundary of the strip, say at
$y=l/L \to 0$, irrespective to the other. This density, denoted by $\rho_0(l/L)$, in the
continuum approximation is calculated for percolation by conformal methods\cite{SKDZ07} as:
\be
\rho_0(y) \propto \left(\sin \pi y\right)^{-x_b} \left(\cos \frac{\pi y}{2} \right)^{x_s}\;
\label{rho_0}
\ee
This density is analogous to the order parameter profile in the system with fixed-free boundary conditions\cite{CFT91a,CFT91b}.
\begin{figure}
  \begin{center}
     \includegraphics[width=3.35in,angle=0]{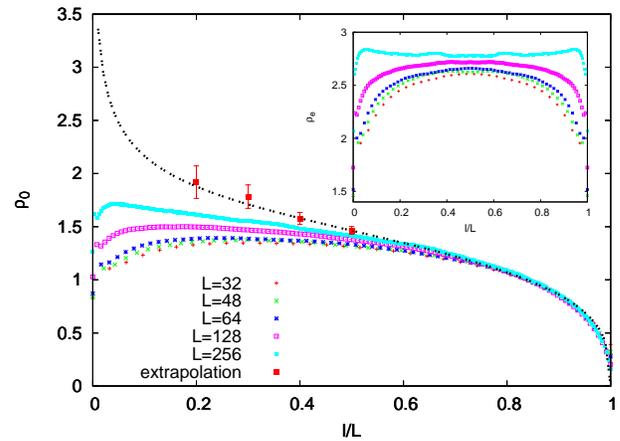}
   \end{center}
   \caption{(Color online) Density profiles, $\rho_0(l/L)$, of the RBPM for different
widths by using such normalization that the curves have the same
asymptotics around $y=1$. The extrapolated values for $l/L \ge 0.5$ are denoted
by red squares.
The dashed line indicates the conformal result in Eq.(\ref{rho_b}) with the conjectured exponents
in Eq.(\ref{x_x_s}). Inset: density profiles $\rho_e(l/L)$ for different
widths.}
   \label{fig:rho_0}
 \end{figure}
The numerically calculated densities are shown in Fig.\ref{fig:rho_0} for different widths, where
we used such a normalization that the curves have the same asymptotics at the free boundary, i.e. around $y=1$.
Close to the free boundary the densities for different $L$ fall to the same curve, which
is well described by the conformal formula in Eq.(\ref{rho_0}) in which
we have used the conjectured  exponents in Eq.(\ref{x_x_s}). In larger distance from the
free surface the numerically calculated densities start to deviate from the conformal result for some $y<\tilde{y}_L$
and $\tilde{y}_L$ is a decreasing function of $L$. If we extrapolate the simulated profiles with $L$ the
region of agreement with the conformal result is extended to the interval $0.2< y \le 1.$, as seen in Fig.\ref{fig:rho_0}.

The finite-size dependence of the densities in this case can be attributed to the effect
of the finite breaking-up length, $l_b$. Close to the touched surface in the continuum limit,
$l_b \ll l \ll L$, the density is described by the scaling result by Fisher and de Gennes\cite{fisher_degennes}: $\rho_0(l) \sim l^{-x_b}$. However by approaching the breaking-up length, $l_b$,
the increase of the profile is stopped and for $l<l_b~~\rho_0(l)$ start to decrease. This is
due to the structure of the connected clusters close to the surface. As seen
in Fig.\ref{fig:cluster} the number of touching sites in a cluster is comperatively smaller
for the RBPM with $\Delta<1/2$ (upper and middle panel of Fig.\ref{fig:cluster}), than
for percolation with $\Delta=1/2$ (lower panel of Fig.\ref{fig:cluster}). Also
for finite widths the small and medium size touching clusters
are rarely represented for the RBPM. By approaching with $l$
the other, free side of the strip the crossing clusters start to bring the dominant contribution to the
density, $\rho_0(l/L)$, which is then well described by the conformal formula.

\begin{figure}
  \begin{center}
     \includegraphics[width=3.35in,angle=0]{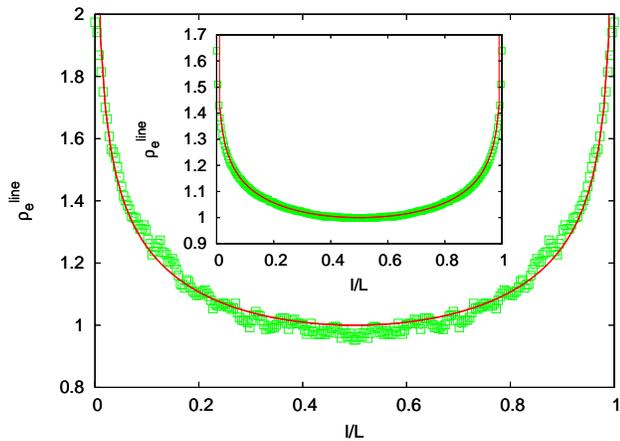}
   \end{center}
   \caption{(Color online) Density profile along a vertical line with two touching boundary
points, $\rho_e^{\rm line}(l/L)$, for the RBPM. The dashed line indicates the conformal result in Eq.(\ref{rho_e}) with the conjectured exponents
in Eq.(\ref{x_x_s}). In the inset the same quantity is shown for percolation. Here in the analytical
formula in Eq.(\ref{rho_e}) we use $x_b=5/48$ and $x_s=1/3$. In both figures the boundary parameter
in Eq.(\ref{y_a}) is $a=0$.
}
   \label{fig:rho_e}
 \end{figure}

Finally we consider $\rho_e(l/L)$, which is the density of points in such clusters which are touching
either the boundary at $l=1$ or at $l=L$ or both. For percolation this density is predicted
through conformal invariance as\cite{SKDZ07}:
\be
\rho_e(y) \propto \left(\sin \pi y\right)^{-x_b}\;
\label{rho_e}
\ee
and it is analogous to the order parameter profile with parallel fixed spin boundary conditions\cite{BE85}.
Note that we have the relation: $\rho_b(y)=\rho_0(y)+\rho_1(y)-\rho_e(y)$, with $\rho_1(y)=\rho_0(1-y)$.
For the RBPM this density is strongly perturbed by the finite breaking-up length
at both boundaries as can be seen in the inset of Fig.\ref{fig:rho_0}. In this case we did not try to
perform an extrapolation and conclude that even larger finite systems would be necessary to test
the conformal predictions in a direct calculation. In order
to try to test the result in Eq.(\ref{rho_e}) we studied another density which is defined on
crossing clusters, so that one expects to be represented correctly in smaller systems, too.
Here we define a density, $\rho^{\rm line}_e(l/L)$, in crossing clusters and consider
points only in such vertical lines, where at both ends of the given
line the cluster touches the boundaries. Since $\rho^{\rm line}_e(l/L)$ is related to the
operator profile with fixed-fixed boundary conditions we expect that it has the same scaling
form as the previously defined density, $\rho_e(l/L)$. In Fig.\ref{fig:rho_e} we show the calculated
densities for the RBPM, which is compared with the analytical prediction in Eq.(\ref{rho_e}). A
similar analysis for percolation is shown in the inset of Fig.\ref{fig:rho_e}. In both cases we found that
the numerical and analytical results for this type of profile are in satisfactory agreement,
although the statistics of the numerical data is somewhat low, since just a fraction of $\sim L^{-2x_s} \sim
L^{-1}$ lines can be used in this analysis.

We can thus conclude that all the critical densities we considered for the RBPM are found in agreement
with the theoretical prediction, which is obtained from the corresponding conformal results for
percolation by replacing the scaling dimensions with the appropriate (conjectured) values for the RBPM.
From an analysis of the profile, $\rho_b(l)$, close to the boundary we have obtained new accurate
estimate of the critical
exponent, $x_s-x_b$, giving further support of the conjecture in Eq.(\ref{x_x_s}).

\section{Final clusters in the RTFIC}
\label{sec:RTFIC}
\subsection{Model}

The random transverse-field Ising chain is defined by the Hamiltonian:
\be
\hat{\mathcal{H}}=-\sum_i J_i \sigma_i^x \sigma_{i+1}^x - \sum_i h_i \sigma_i^z\;
\label{hamilton_I}
\ee
in terms of the Pauli matrices, $\sigma_i^{x,z}$, at site $i$. The couplings, $J_i$, and the
transverse fields, $h_i$, are independent and identically distributed random numbers. The
critical point of the system is located at: $[\ln h]_{\rm av}=[\ln J]_{\rm av}$, where we
use the notation $[\dots]_{\rm av}$ to indicate the average over quenched disorder.

We note that the RTFIC is the Hamiltonian version\cite{kogut} of the McCoy-Wu model\cite{mccoywu},
which is a 2d Ising model with layered disorder. In the $i$-th layer of this model the couplings in the
vertical and horizontal directions are given by: $K_1(i)$ and $K_2(i)$, respectively, which are
related to the parameters of the RTFIC as: $h_i=-\tau^{-1} \tanh^{-1}\exp(-2K_1(i))$ and
$J_i=-\tau^{-1} K_2(i)$, where in the Hamiltonian limit $\tau \to 0$.

\subsection{SDRG method}

The RTFIC can be efficiently studied within the frame of a renormalization group approach\cite{mdh,im},
which is expected to lead to asymptotically exact results\cite{fisher}. The basic feature of this
procedure is to successively eliminate those degrees of freedom which have the largest
local energy scale and thus represents the fastest local mode. At a given step of the
renormalization the global energy scale is defined by $\Omega=\max\{J_i,h_i\}$ and the
local term of value $\Omega$ is eliminated from the Hamiltonian. Here we have two different elementary renormalization steps: cluster formation and
cluster decimation.

{\it i) Cluster formation}: if the largest local parameter is a coupling, say $J_2 \gg h_2,h_3$
($h_2$ and $h_3$ being the transverse fields acting at the two ends of $J_2$)
then a new spin cluster is formed in an effective transverse field: 
$\tilde h_{23} \approx h_2 h_3 /J_2$, which is calculated in second-order perturbational calculation.
The moment of the new cluster is given by $\tilde \mu_{23}=\mu_{2}+\mu_{3}$, in terms of the
moments of the original clusters, $\mu_{2}$ and $\mu_{3}$. In the starting Hamiltonian all spins
have the same moment of unity.

{\it ii) Cluster decimation}: if the largest local parameter is a transverse field,
say $h_2 \gg J_2,J_3$ ($J_2$ and $J_3$ being the couplings which are connected to the
site with $h_2$) then the spin cluster is decimated out and an effective coupling
$\tilde J_{23} \approx J_2 J_3 /h_2$ is formed between the remaining sites. If the decimated
spin is at the boundary of an open chain no new couplings are formed.

During renormalization we repeat the elementary decimation steps, which at the starting period
are only approximative, but as the energy scale is reduced and
the fixed point, $\Omega^*=0$, is approached they become asymptotically exact. In this
limit the renormalization group equations can be solved
analytically. The length-scale of the clusters (and bonds), defined by the linear size of
the original region which is renormalized to the new variable is shown\cite{fisher} to scale as
\be
\ell \sim \ln \left( \Omega/\Omega_0 \right)^2\;,
\ee
where $\Omega_0$ is a reference energy scale. On the other hand the average cluster moment
behaves as\cite{fisher}:
\be
\mu \sim \ln \left( \Omega/\Omega_0 \right)^{\Phi},\quad \Phi=\frac{1+\sqrt{5}}{2}\;.
\ee
Note that the average magnetization at the critical point behaves as: $m(\ell)\sim \mu/\ell\sim  \ell^{-x_b}$
as lengths are rescaled by a factor, $\ell$, and $x_b=1-\Phi/2$ is just the scaling dimension
introduced in Eq.(\ref{x_x_s}).

\subsection{Densities of critical clusters}

\begin{figure}
  \begin{center}
    \includegraphics[width=3.45in,angle=0]{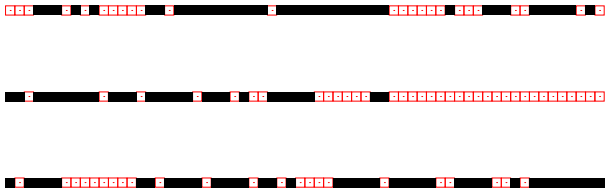}
   \end{center}
   \caption{(Color online) Examples of final clusters in the RTFIC for $L=64$.
Sites in the final cluster are denoted by black squares. Upper panel: final cluster in
general position; middle panel: final cluster containing the boundary spin at $\ell=1$;
lower panel: final cluster containing both boundary spins at $\ell=1$ and at $\ell=L$.
}
   \label{fig:final}
 \end{figure}

Having a finite chain with length, $L$, we perform the decimation until the final cluster, which have
a moment of $\mu(L) \sim L^{\Phi/2}$. Sites of the original chain which belong to the final
cluster are very strongly correlated and we can ask questions about the density of sites
in the final cluster, i.e. about the probability that a given site is contained in a
final cluster. The structure of spins in the final clusters are illustrated in Fig.\ref{fig:final}.
Note that the final cluster in the 1d space is disconnected, and the correlations are realized
along the (imaginary) time direction.
Densities of critical clusters in the RTFIC is studied numerically. We have considered a large number of
($3 \times 10^7$) chains of length $L=2^{13}=8192$ with open boundary conditions. We used
the same type of uniform disorder: $p(u)=1$, for $0 \le u \le 1$ and $p(u)=0$, for $u>1$, both for the
couplings and the transverse fields, in this way we have satisfied the criticality condition.
The strong disorder renormalization procedure is performed for each chain up to the final
spin cluster and then the statistics of the sites belonging to the final clusters are investigated.

We have studied the density of three different class of clusters, which have somewhat
analogous definitions to the clusters studied for the RBPM. In terms of all final clusters (see
the upper panel of Fig.\ref{fig:final}) we define $\hat{\rho}(l/L)$. If we consider those
final clusters which have the boundary point $l=1$ (see the middle panel of Fig.\ref{fig:final})
we obtain $\hat{\rho}_0(l/L)$. Finally, if the clusters contain both boundary points,
$l=1$ and $l=L$ (see the lower panel of Fig.\ref{fig:final}) we define $\hat{\rho}_{01}(l/L)$.
We note that for these densities no analytical conjecture is available, since the system is
not conformally invariant.
\begin{figure}
  \begin{center}
     \includegraphics[width=3.35in,angle=0]{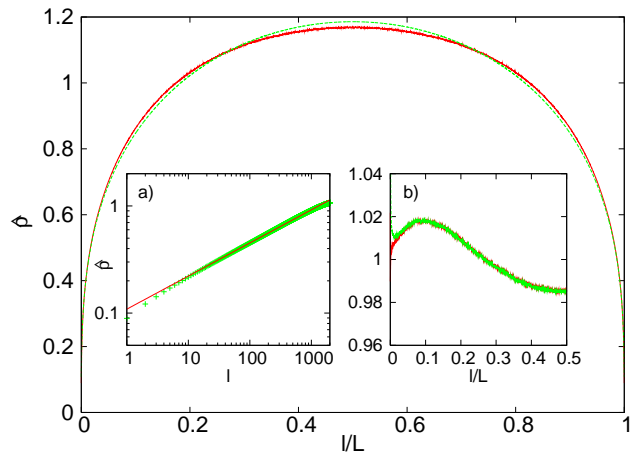}
   \end{center}
   \caption{(Color online)Density profile of the RTFIC considering all final clusters of
the SDRG procedure. The dashed (red) line indicates the formula in Eq.(\ref{rho_b}).
Insets: a) Density profile close to the surface in log-log plot. The straight (red) line
has a slope $x_s-x_b=0.309$. b) Ratio of the numerical results and the formula in Eq.(\ref{rho_b}).
The parameter in Eq.(\ref{y_a}) is a=0. (full or red line) and a=-1. (dashed or green line).
}
   \label{fig:rho_hat_a}
 \end{figure}

The density of all final clusters is shown in
Fig.\ref{fig:rho_hat_a}. First we note that close to the boundaries
the behavior of the profile is predicted by scaling theory as $\hat{\rho}(y)
\sim y^{x_s-x_b},\quad y \ll 1$, or $\hat{\rho}(\ell)
\sim \ell^{x_s-x_b},\quad \ell \ll L$. This relation is indeed satisfied as shown in inset a) of Fig.\ref{fig:rho_hat_a}. From this inset we can notice that the microscopic length-scale
of the model, $l_{m}$, is just a few lattice spacing and for $l > l_m$ the calculated
profile is well described by the asymptotic scaling result. The scaling result
is valid for the formula in Eq.(\ref{rho_b}) with the appropriate scaling dimensions, therefore we tried to
compare it with the numerical results. As seen in Fig.\ref{fig:rho_hat_a} the agreement is very
good for all values of $y$. To have a more precise check in inset a) of Fig.\ref{fig:rho_hat_a}
we have presented the ratio of the numerical results and
the formula in Eq.(\ref{rho_b}). Here one can notice small deviations from unity, which are of the
order of 1\%. Consequently the formula in Eq.(\ref{rho_b}) is a very good fit, however presumably it
is not exact.

We can conclude that the density profiles of the final clusters of the SDRG procedure for the RTFIC
are well described by scaling predictions close to the boundaries. The full profiles are also well
approximated by analytical formulae, which are however not fully perfect.

\begin{figure}
  \begin{center}
     \includegraphics[width=3.35in,angle=0]{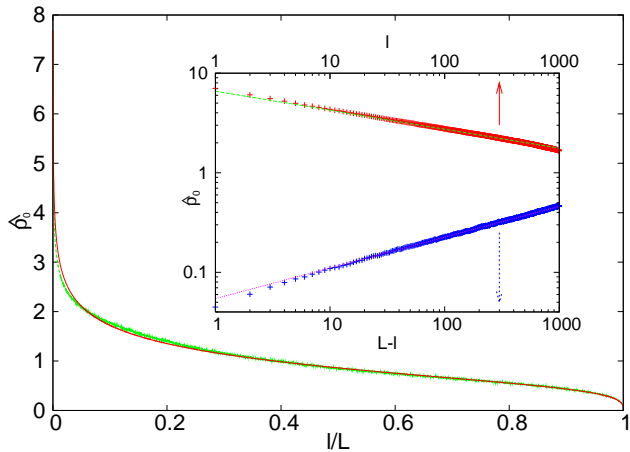}
   \end{center}
   \caption{(Color online) Density profile of the RTFIC considering those final clusters of
the SDRG procedure which contain the site at $l=1$. The dashed line indicates the formula in Eq.(\ref{rho_hat_b}). Inset: density profiles at the two surfaces in log-log plot.
The straight lines have slopes: $-x_b=-0.191$ (upper or green) and $x_s-x_b=0.309$ (lower or red).
}
   \label{fig:rho_hat_b}
 \end{figure}

The density of final clusters which contain the boundary site at $l=1$ is shown in Fig.\ref{fig:rho_hat_b}.
From scaling theory one knows the behavior of the profile close to the boundaries:
$\hat{\rho}_0(y)\sim (y)^{-x_b},\quad y \ll 1$ (or $\hat{\rho}_0(\ell)\sim (\ell)^{-x_b},\quad \ell \ll L$) and $\hat{\rho}_0(y)\sim (1-y)^{x_s-x_b},\quad 1-y \ll 1$ (or $\hat{\rho}_0(L-\ell)\sim (L-\ell)^{x_s-x_b},
\quad L-\ell \ll L$), respectively. This behavior is indeed found in the numerically calculated profile
as seen in the inset of Fig.\ref{fig:rho_hat_b}.
The asymptotics mentioned above is valid for the formula in Eq.(\ref{rho_0}). We tried to fit the
numerical results with this formula (with the appropriate scaling dimensions), however the weight of
the the tail at $y \sim 1$ given by this formula is too large, by about a factor
of 2. Much better agreement with the data can be obtained with the formula:
\be
\hat{\rho}_0(y) \propto \left(\sin \pi y\right)^{-x_b} \left[\left(\cos \frac{\pi y}{2} \right)^{x_s}-
\left(\sin \frac{\pi y}{2} \right)^{x_s}+1\right]\;,
\label{rho_hat_b}
\ee
which is just the average of the density of clusters which touch one boundary an may and may not touch
the other boundary. As seen in Fig.\ref{fig:rho_hat_b} the analytical and numerical results are
close to each other for all $y$, although the agreement is certainly not perfect.

\begin{figure}
  \begin{center}
     \includegraphics[width=3.35in,angle=0]{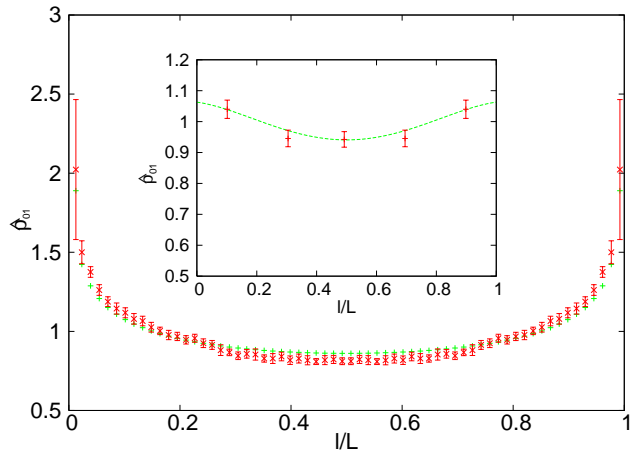}
   \end{center}
   \caption{(Color online) Density profile of the RTFIC considering those final clusters of
the SDRG procedure which contain both boundary sites. The dashed line indicates the formula
in Eq.(\ref{rho_e}). Inset: ratio of the numerical results and the formula in Eq.(\ref{rho_e}).
Typical errorbars are also indicated.}
   \label{fig:rho_hat_c}
 \end{figure}

Finally we consider those final clusters that touch both boundaries. The corresponding
density, $\hat{\rho}_{01}(y)$, is similar to the order parameter profile with fixed-fixed boundary
condition and its functional form for percolation is given in Eq.(\ref{rho_e}). The numerically
calculated profile is given in Fig.\ref{fig:rho_hat_c}. Here the comperatively large fluctuations of
the data points are due to the fact that only a fraction $\sim L^{-2x_s} \sim L^{-1}$ of the
samples have a final cluster which touches both boundaries. We have compared the calculated profile
with the analytical formula in Eq.(\ref{rho_e}) using $x_b$ from Eq.(\ref{x_x_s}). The agreement is
generally very good, but not fully perfect. Small deviations of the order of a few percents can
be observed (see in the inset of Fig.\ref{fig:rho_hat_c}).

\section{Discussion}
\label{sec:Disc}

In this paper we have studied the density of critical clusters in two models the critical properties
of which are dominated by disorder effects. One of the models is the two-dimensional random bond
Potts model and we considered the FK clusters in the large-$q$ limit. This model is expected to be
conformally invariant, which means that average
quantities which are related to FK clusters (such as correlation function and
magnetization densities) are invariant under conformal transformations. We note that
the RBPM and conventional bond percolation represent two different fixed points of the same
phase diagram, which correspond to $0<\Delta<1/2$ and $\Delta=1/2$ for the binary
disorder, respectively, see Fig.\ref{fig:cluster}. In contrast to percolation in the RBPM there is a finite
length-scale, the breaking-up length, $l_b$, and results of the continuum approximation are
expected to hold for lengths which are larger than $l_b$. In the strip geometry
we have calculated the density of points of different type of clusters (crossing clusters, clusters
which touch one boundary of the strip, etc.) in analogy with a related study of percolation
in\cite{SKDZ07}. The densities close to free surfaces are well described by scaling predictions
and from this analysis accurate estimate of the critical exponent $x_s-x_b$ is obtained in
agreement with the conjecture in Eq.(\ref{x_x_s}). The full profiles are compared with analytical formulae
which are obtained from the corresponding conformal results for percolation by using the appropriate
values of the bulk and surface scaling exponents in Eq.(\ref{x_x_s}). We have
observed that the numerically calculated profiles agree well with the conformal results
outside the surface region of width $\sim l_b$.

The second model we considered is the RTFIC, the fixed point of which is expected to control the
critical behavior of a large class of 2d classical systems with anisotropic randomness. Examples are
the Ising model and the (directed) percolation with layered disorder. In these systems scaling
at the critical point is strongly anisotropic, therefore these systems are not conformally invariant.
In the RTFIC critical clusters are defined through the strong disorder RG procedure. Here
spins in the final cluster are strongly
correlated and play analogous role as clusters in percolation or FK clusters in the Potts model. The
density of final clusters of the RTFIC close to the surfaces of the strip are shown to obey scaling
relations. We also tried to find analytical formulae which correctly approximate the numerical profiles.
These formulae, which are borrowed from similar studies of conformal systems have an overall very
good description, however these are not fully perfect. We have noticed a discrepancy of the order of
a few percent.

Our investigations can be extended into different directions. For 2d classical systems one can study
the density of FK clusters in the $q$-state Potts model, both without disorder (for $q \le 4$)
and in the presence of disorder (for general value of $q$). One can also study the density of
geometrical clusters in the 2d random-field Ising model\cite{sepp,ki07}. For the random transverse-field
Ising model one possibility is to investigate the distribution of final clusters in a 2d strip.

We thank for useful discussions with L. Turban.
This work has been supported by the National Office of Research and Technology under
Grant No. ASEP1111, by German-Hungarian exchange programs (DAAD-M\"OB and
DFG-MTA) and by the
Hungarian National Research Fund under grant No OTKA TO48721, K62588, MO45596.
M.K. thanks the Minist\`ere Fran\c{c}ais des
Affaires \'Etrang\`eres for a research grant.

\end{document}